\documentclass[12pt]{article}
\usepackage{amsmath,amssymb,amsthm,amsxtra,overpic,bbm,bm,epsfig,subfigure}
\usepackage{hyperref}
\usepackage{color}
\usepackage{stackengine} % adjust row height in table
\usepackage{multirow}
\usepackage{slashed}
\usepackage{booktabs} % \toprule etc

\begin{document}
\vspace{0.2cm}

\begin{center}
{\Large\bf Leptogenesis from the Dirac CP-violating phase in the minimal left-right symmetric model}\\
\vspace{0.2cm}
%\today
\end{center}
\vspace{0.2cm}

\begin{center}
{\bf Xueke Chen}~$^{a}$,~{\bf Xinyi Zhang}~$^{a,b,c,d}$~\footnote{Email: zhangxy@hbu.edu.cn}
\\
\vspace{0.2cm}
$^a$ {\em \small Department of Physics, Hebei University, Baoding, 071002, China}\\
$^b$ {\em \small Hebei Key Laboratory of High-precision Computation and Application of Quantum Field Theory, Baoding, 071002, China}\\
$^c$ {\em \small Hebei Research Center of the Basic Discipline for Computational Physics, Baoding, 071002, China}\\
$^d$ {\em \small SISSA, International School for Advanced Studies, Via Bonomea 265, 34136 Trieste, Italy}
\end{center}

\vspace{0.5cm} 
 
\begin{abstract}
Leptogenesis from low-energy CP violation provides a vital link between neutrino physics and the observed baryon asymmetry of the universe. However, this connection is typically obscured by unknown high-energy parameters. In this work, we investigate thermal leptogenesis in the Minimal Left-Right Symmetric Model with generalized parity as the left-right symmetry, where the hermiticity of the Dirac neutrino coupling allows the right-handed mixing matrix $V_\mathrm{R}$ to be determined with minimal assumptions. We show that for a real $V_\mathrm{R}$, these conditions favor CP-conserving Majorana phases, leaving the Dirac CP-violating phase ($\delta$) as the sole source of asymmetry. By numerically exploring all four leptogenesis scenarios, we demonstrate that $\delta$ alone can generate the observed baryon asymmetry with the correct sign within specific regions of the parameter space. The results exhibit a high sensitivity to the neutrino mass ordering and the lightest neutrino mass, providing a stringent, testable framework for future experimental measurements of the CP phase and neutrino mass scale.
\end{abstract}

\section{Introduction}
Neutrino physics has entered the era of precision measurement. With mixing angles now determined within percent-level uncertainty, the next critical objectives are to resolve the neutrino mass ordering and measure the magnitude of Dirac CP violation~\cite{JUNO:2015zny,DUNE:2015lol,Hyper-Kamiokande:2018ofw}. A pressing question in this field is the extent to which low-energy CP violation contributes to the Baryon Asymmetry of the Universe (BAU) via leptogenesis—an elegant and compelling framework for explaining the origin of matter. Exploring this potential link remains a primary frontier at the intersection of particle physics and cosmology. The possibility that the BAU can be accounted for by low-energy CP violation via leptogenesis is particularly intriguing, as it simultaneously addresses the origin of beyond-the-Standard-Model neutrino masses. Furthermore, since traditional seesaw mechanisms for neutrino mass generation typically involve high-energy scales beyond the reach of current colliders, establishing such a connection may provide a rare and vital test for an otherwise inaccessible sector of physics.

The connection between low-energy CP violation and leptogenesis is not straightforward. The neutrino Yukawa coupling—or equivalently the neutrino Dirac mass—governs the CP asymmetry generated by right-handed neutrino (RHN) decays. This coupling receives contributions from both a high-energy sector associated with the heavy RHNs and a low-energy sector associated with the light left-handed neutrinos. In principle, both sectors can contain CP-violating phases that contribute to the final baryon asymmetry through leptogenesis. To separate these contributions, one can use the Casas-Ibarra parametrization~\cite{Casas:2001sr}, which expresses the neutrino Dirac mass in the fermion mass basis as:
\begin{align}
M_\mathrm{D} = i D_N^{1/2} R D_\nu^{1/2} V_\mathrm{L}^\dagger, \label{eq:CI}
\end{align}
where $D_\nu$ and $D_N$ denote the diagonal mass matrices for light and heavy neutrinos, respectively, and $V_\mathrm{L}$ corresponds to the light neutrino mixing matrix. The $R$ matrix is a complex orthogonal matrix that contains potential CP violation from the high-energy sector and parametrizes our ignorance of the neutrino Yukawa coupling. Even if one excludes the specific scenario where contributions from these two sectors cancel out, the unknown high-energy phases can significantly affect the results, limiting our ability to make a definite prediction of the baryon asymmetry arising solely from low-energy CP violation.

Seeking a potential connection between low-energy CP violation and leptogenesis has been a subject of intense investigation for over two decades (e.g., \cite{Branco:2001pq, Davidson:2002em, Frampton:2002qc, Endoh:2002wm, Branco:2002xf, Pascoli:2003uh}). Existing methods used to identify a possible connection include minimal lepton flavor violation~\cite{Branco:2006hz,Merlo:2018rin}, flavor symmetries~\cite{Hagedorn:2009jy,Meroni:2012ze,Karmakar:2014dva,Gehrlein:2015dxa,Ishihara:2015uua,Li:2017zmk,Chauhan:2021xus,Drewes:2022kap}, generalized CP symmetry~\cite{Chen:2016ptr,Hagedorn:2016lva,Li:2017zmk}, and quantum corrections~\cite{Xing:2020erm,Xing:2020ghj}.
A general connection was first identified in the flavored regime \cite{Pascoli:2006ie, Pascoli:2006ci}, the flavor effect is found to have a non-negligible effect even at extremely high temperatures ($T \gg 10^{12}$ GeV)~\cite{Moffat:2018smo}.  Recently, density matrix equations were used to correlate the baryon asymmetry sign with $\sin\delta$~\cite{Granelli:2021fyc}. Furthermore, viable leptogenesis may arise purely from the Dirac CP-violating phase in low-scale freeze-in models~\cite{Granelli:2023tcj}.

The Minimal Left-Right Symmetric Model (MLRSM)~\cite{Pati:1974yy,Mohapatra:1974gc,Senjanovic:1975rk,Senjanovic:1978ev} provides a unique framework for pinning down this connection. As pointed out in Refs.~\cite{Nemevsek:2012iq,Senjanovic:2016vxw,Senjanovic:2018xtu,Senjanovic:2019moe}, the neutrino Yukawa couplings can be completely determined by light and heavy neutrino masses and mixings in the MLRSM when either generalized parity ($\mathcal{P}$) or charge conjugation ($\mathcal{C}$) acts as the LR symmetry. This determination eliminates the $R$-matrix uncertainty, thus providing an excellent pathway for connecting low-energy CP violation to leptogenesis. Furthermore, the MLRSM features rich phenomena concerning neutrinos and leptogenesis. Both Type I and Type II seesaw mechanisms are realized naturally; regarding leptogenesis, the decaying particle can be a triplet scalar in addition to the RHNs. Numerous successful realizations of leptogenesis exist within left-right symmetric theories, focusing on different regimes and methods to constrain the free parameters of the neutrino Yukawa sector~\cite{Joshipura:2001ya,Rodejohann:2002mh,Babu:2005bh,Akhmedov:2006yp,Chao:2007rm,Hallgren:2007nq,Abada:2008gs,BhupalDev:2014hro,Gu:2014mga,BhupalDev:2019ljp,Rink:2020uvt,Zhang:2020lir,Patel:2023voj,Babu:2024glr,Kumar:2025bfe}.

We investigate the intriguing possibility that the Dirac CP-violating phase is the sole source of CP violation required for unflavored thermal leptogenesis in the MLRSM. The same subject was investigated in Ref.~\cite{Zhang:2020lir} with $V_\mathrm{R}=V_\mathrm{L}$. In this work, instead of making assumptions about $V_\mathrm{R}$, we use the hermiticity of $M_\mathrm{D}$ to determine $V_\mathrm{R}$. Although this improvement is technically involved, it provides a truly consistent map of the parameter space to the final baryon asymmetry, establishing a direct connection to the Dirac CP phase with minimal assumptions.

This paper is organized as follows. In Sec.~\ref{sec:framework}, we set up the basic framework for our discussion. In Sec.~\ref{sec:conditions}, we detail how conditions for the $V_\mathrm{R}$ parameters arise from the model structure and how the free parameters can be minimized in a meaningful way. In Sec.~\ref{sec:results}, we present and discuss the numerical results. Finally, we conclude in Sec.~\ref{sec:conclusions}.

\section{The framework}\label{sec:framework}
The part of the MLRSM Lagrangian relevant to our discussion is
\begin{align}
\mathcal{L} \supset -\bar{l}_\mathrm{L}^{} \left(Y_1 \Phi_1 - Y_2^{} \Phi_2^* \right) l_\mathrm{R}^{} - \frac{1}{2} \left(l_\mathrm{L}^T\mathrm{C} Y_\mathrm{L}^{} i \sigma_2^{} \Delta_\mathrm{L}^{} l_\mathrm{L}^{}
 + l_\mathrm{R}^T  \mathrm{C}  Y_\mathrm{R}^{} i \sigma_2^{} \Delta_\mathrm{R}^{} l_\mathrm{R}^{}\right) -\lambda_{ij} \mathrm{Tr} \left( \Delta_\mathrm{R}^\dagger \Phi_i^{} \Delta_\mathrm{L}^{} \Phi_j^\dagger \right)+ \mathrm{h.c.},\label{eq:Lag}
\end{align}
where $i,j=1,2$ and
\begin{align}
l_{\mathrm{L,R}}=\left(\begin{array}{c}
\nu\\ e\\
\end{array}\right)_\mathrm{L,R},~~
\Delta_\mathrm{L,R}=\left(\begin{array}{cc}
\delta^+/\sqrt{2}  & \delta^{++}\\
\delta^0 & -\delta^+/\sqrt{2}\\
\end{array}\right)_\mathrm{L,R},~~
\Phi_1= \left( \begin{array}{cc}
\phi_1^0 & \phi_2^+\\
\phi_1^- & -\phi_2^0\\
\end{array}\right),~~
\Phi_2=\sigma_2^{} \Phi_1^* \sigma_2^{}.
\end{align}
The fields and their representations under the MLRSM gauge group $SU(2)_\mathrm{L} \times SU(2)_\mathrm{R} \times U(1)_{\mathrm{B}-\mathrm{L}}$ are shown in Table \ref{tab:assignment}. In addition to these gauge symmetries, the MLRSM features a discrete left-right symmetry—either $\mathcal{P}$ or $\mathcal{C}$—both of which enforce the equality of the gauge couplings, $g_\mathrm{L} = g_\mathrm{R} = g$.

The MLRSM gauge symmetry is broken down to the SM gauge symmetry when the neutral component of $\Delta_\mathrm{R}$ obtains a vacuum expectation value (VEV), $\langle \delta_\mathrm{R}^0 \rangle = v_\mathrm{R}$. Subsequently, the electroweak symmetry breaks via $\langle \phi^0_i \rangle = v_i$ (for $i=1, 2$), which triggers an induced symmetry breaking for $\Delta_\mathrm{L}$ with $\langle \delta_\mathrm{L}^0 \rangle = v_\mathrm{L}$. Following the completion of all symmetry-breaking stages, the Lagrangian in Eq.~\eqref{eq:Lag} leads to the following neutrino mass terms:
\begin{align}
    \mathcal{L}_\nu \supset -\dfrac{1}{2} (\overline{ \nu_\mathrm{L}^c}, \overline{N_\mathrm{L}^c}) 
\mathcal{M} 
    \begin{pmatrix}
        \nu_\mathrm{L} \\
        N_\mathrm{L}
    \end{pmatrix}
    + {\rm h.c.}\;,
\end{align}
where we define $N_\mathrm{L} \equiv C \bar{\nu}_\mathrm{R}$, assuming the hierarchy $v_\mathrm{R} \gg v \gg v_\mathrm{L}$. The resulting $6 \times 6$ effective neutrino mass matrix is:
\begin{align}
\mathcal{M}  =
\begin{pmatrix}
M_\mathrm{L} & M_\mathrm{D}^T\\
M_\mathrm{D}& M_N
\end{pmatrix}\;,\label{eq:fullM}
\end{align}
with $M_\mathrm{L} = Y_\mathrm{L} v_\mathrm{L}$, $M_N = Y_\mathrm{R}^* v_\mathrm{R}$, and $M_\mathrm{D} = Y_1 v_1 - Y_2 v_2 $.
The full $6 \times 6$ neutrino mass matrix can be block-diagonalized, from which we obtain the light neutrino mass matrix:
\begin{align}
\label{eq:Mnu}
M_\nu = M_\mathrm{L} - M_\mathrm{D}^T \dfrac{1}{M_N} M_\mathrm{D} \equiv M_\nu^{\mathrm{II}}+M_\nu^{\mathrm{I}}\;.
\end{align}
%\dfrac{v_\mathrm{L}}{v_\mathrm{R}}M_N^*
Generally, the light neutrino mass matrix in the MLRSM is composed of both Type-I and Type-II seesaw contributions.

\begin{table}[t]
\centering
\caption{\label{tab:assignment}Field representations under the gauge groups in MLRSM.}
\vspace{.2cm}
\begin{tabular}{l|cccccc}
\toprule \hline
~ & $l_L$ & $l_R$ & $\Delta_L$ & $\Delta_R$ & $\Phi_1$ & $\Phi_2$ \\
\midrule
$SU(2)_L$ & $2$ & $1$ & $3 $ & $1 $ & $2 $ & $2$ \\
$SU(2)_R$ & $1$ & $2$ & $1$ & $3$ & $2$ & $2$ \\
$U(1)_{B-L}$ & $-1$ & $-1$ & $2$ & $2$ & $0$ & $0$ \\
\hline
\bottomrule
\end{tabular}
\end{table}

\subsection{$M_\mathrm{D}$ in MLRSM}
The as-yet-unobserved heavy neutrinos interact with SM fields through Yukawa interactions, making these couplings a primary target for experimental probes. More importantly, these interactions contribute to light neutrino masses, which remain at the frontier of current experimental research. It has been pointed out that in the MLRSM, Yukawa couplings are completely determined by light and heavy neutrino masses and mixings—provided that either $\mathcal{C}$ or $\mathcal{P}$ acts as the LR symmetry and remains unbroken by the VEV of $\Phi$~\cite{Nemevsek:2012iq,Senjanovic:2016vxw,Senjanovic:2018xtu,Senjanovic:2019moe}. We briefly review these findings below.

Under the $\mathcal{P}$ transformation, the fields transform as:
\begin{align}
l_\mathrm{L} \leftrightarrow l_\mathrm{R},~~\Delta_\mathrm{L} \leftrightarrow \Delta_\mathrm{R},~~\Phi^{} \leftrightarrow \Phi^\dagger.
\end{align}
Such transformation leads to
$Y_{1,2}^{}=Y_{1,2}^\dagger,~~Y_\mathrm{L}=Y_\mathrm{R} \equiv Y_\mathrm{T}$
and further results in 
\begin{align}
M_\mathrm{D}=M_\mathrm{D}^\dagger
\end{align}
in the case of unbroken parity. 

To see how $M_\mathrm{D}$ can be expressed in terms of light and heavy neutrino masses and mixings, one first writes down the effective light neutrino mass matrix \eqref{eq:Mnu} in this scenario:
\begin{align}
M_\nu = \frac{v_\mathrm{L}}{v_\mathrm{R}}  M_N^* - M_\mathrm{D}^T \frac{1}{M_N}M_\mathrm{D}\,.
\end{align}
Introducing a matrix $H$ satisfying
\begin{align}
H^{} H^T \equiv \frac{v_\mathrm{L}^*}{v_\mathrm{R}} - \frac{1}{\sqrt{M_N}} M_\nu^* \frac{1}{\sqrt{M_N}} \,, \label{eq:HHT}
\end{align}
$M_\mathrm{D}$ can be written into a compact form
\begin{align}
M_\mathrm{D} =\sqrt{M_N} H \sqrt{M_N^*}\,, \label{eq:MDH}
\end{align}
where $H$ is Hermitian, following the hermiticity of $M_\mathrm{D}$. It is clear from Eqs.~\eqref{eq:HHT} and \eqref{eq:MDH} that $M_\mathrm{D}$ is uniquely determined once the light and heavy neutrino mass matrices are known. Applying these relations to the Type-I seesaw limit (i.e., $M_\nu \simeq M_\nu^\mathrm{I}$), one finds:
\begin{align}
M_\mathrm{D} =i \sqrt{M_\nu^* M_N^*} \,. \label{eq:MDH1}
\end{align}

Under the $\mathcal{C}$ transformation, the fields transform as: 
\begin{align}
l_\mathrm{L}^{} \leftrightarrow l_\mathrm{R}^\mathrm{c},~~\Delta_\mathrm{L}^{} \leftrightarrow \Delta_\mathrm{R}^*,~~\Phi^{}_{} \leftrightarrow \Phi^T_{}\,,
\end{align}
which leads to $Y_{1,2}^{}=Y_{1,2}^T,~~Y_\mathrm{L}^{}=Y_\mathrm{R}^* \equiv Y_\mathrm{T}$ and consequently
\begin{align}
M_\mathrm{D}^{} = M_\mathrm{D}^T\,.
\end{align}
The light neutrino mass matrix is now
\begin{align}
M_\nu = \frac{v_\mathrm{L}}{v_\mathrm{R}}  M_N  - M_\mathrm{D}^T\frac{1}{M_N}M_\mathrm{D}^{} \,.
\end{align}
Following a similar procedure and exploiting the symmetric nature of $M_\mathrm{D}$ in this case, one obtains:
\begin{align}
M_\mathrm{D}=M_N \sqrt{\frac{v_\mathrm{L}}{v_\mathrm{R}}-\frac{1}{M_N}M_\nu}\,.\label{eq:MD_C}
\end{align}
The Type-I seesaw limit also results in a compact expression for the Dirac neutrino mass matrix:
\begin{align}
M_\mathrm{D}= i \sqrt{M_N M_\nu}\,.\label{eq:MDC1}
\end{align}

\subsection{Leptogenesis in MLRSM}
MLRSM provides a rich variety of phenomena concerning leptogenesis. In addition to the fact that both Type I and Type II seesaw mechanisms can be naturally realized, the decaying particle responsible for generating lepton number asymmetry can be the left-handed triplet as well as the RHNs. Consequently, there are four possible scenarios considering unflavored thermal leptogenesis, which we list in Table \ref{tab:class}.

\begin{table}[t!]
 \caption{\label{tab:class} Four possible scenarios of leptogenesis in MLRSM.}\vspace{.2cm}
 \centering
  \begin{tabular}{c|c|c}
   \toprule \hline
  &Lepton asymmetry & Neutrino mass \\ \hline
  Scenario $1$ &  $m_{N_k}<m_\Delta$ & $M_\nu^{\mathrm{I}}>M_\nu^{\mathrm{II}}$ \\ 
   Scenario $2$ &  $m_{N_k}<m_\Delta$ & $M_\nu^{\mathrm{I}}<M_\nu^{\mathrm{II}}$\\
   
   Scenario $3$ &  $m_{N_k}>m_\Delta$ & $M_\nu^{\mathrm{I}}>M_\nu^{\mathrm{II}}$\\
   
   Scenario $4$ &  $m_{N_k}>m_\Delta$ & $M_\nu^{\mathrm{I}}<M_\nu^{\mathrm{II}}$\\
   \hline
  \bottomrule
\end{tabular}
\end{table}

In Scenarios 1 and 2, RHNs are the decaying particles responsible for generating the CP asymmetry. In addition to the standard loop diagrams, the MLRSM includes a supplemental diagram featuring a triplet in the loop, which further contributes to the CP asymmetry \cite{Hambye:2003ka}
\begin{align}
\epsilon_{N_k}^\Delta =\displaystyle -\frac{1}{2 \pi} 
\frac{\sum_{il} \mathrm{Im}\left[(Y_N)_{ki} (Y_N)_{kl} (Y_\mathrm{T}^*)_{il} \mu\right]}{\sum_i |(Y_N)_{ki}|^2 m_{N_k}} 
\left[ 1- \frac{m_\Delta^2}{m_{N_k}^2} \mathrm{ln}\left(1+\frac{m_{N_k}^2}{m_\Delta^2}\right)  \right]\,.\label{eq:epsilonND}
\end{align}
Note that $\epsilon_{N_k}^\Delta$ does not have the usual $Y_N Y_N^\dagger$ dependence found in the conventional CP asymmetry $\epsilon_{N_k}$. Consequently, the left-handed mixing is not cancelled out, allowing low-energy CP violation to persist even in the unflavored regime ($T > 10^{12}$ GeV).
In this regime, the Boltzmann equations governing the evolution are
\begin{align}
\frac{dY_{N_i}}{dz} &= \frac{-z}{s H} 
\left( \gamma_\mathrm{D}^i + \gamma_{\mathrm{S},\Delta \mathrm{L}=1}^i \right) \left( \frac{Y_{N_i}}{Y_{N_i}^{\rm eq}} -1 \right); \nonumber\\
\frac{dY_\Delta}{dz} &=  \frac{-z}{s H} 
\left[ \sum_{i=1}^3 \epsilon_i \left( \gamma_\mathrm{D}^i + \gamma_{\mathrm{S},\Delta \mathrm{L}=1}^i \right) \left( \frac{Y_{N_i}}{Y_{N_i}^{\rm eq}} -1 \right)
- \left(\frac{ \gamma_\mathrm{D}^i}{2} + \gamma_{\mathrm{W},\Delta \mathrm{L}=1}^i \right) \frac{Y_\Delta}{Y_l^{\rm eq}}
\right],\label{eq:NBE}
\end{align}
where $Y_{N_i}$ is the $N_i$ abundance ($Y_{N_i}^{\rm eq}$ is the equilibrium abundance), $Y_\Delta \equiv Y_{\Delta \mathrm{B}}/3 - Y_{\Delta \mathrm{L}}$, and $\epsilon_i \equiv \epsilon_{N_i} + \epsilon_{N_i}^\Delta$. Here, the various $\gamma$ denote the reaction rates (or rate densities) with subscripts indicating the respective processes.

In Scenarios 3 and 4, the triplet is lighter than the RHNs, and the CP asymmetry is generated through triplet decay as:
\begin{align}
\epsilon_\Delta =\displaystyle\frac{1}{8\pi} \sum_k m_{N_k} \frac{\sum_{il} \mathrm{Im} \left[ (Y_N^*)_{ki} (Y_N^*)_{kl} (Y_\mathrm{T})_{il} \mu^* \right] }{\sum_{ij}|(Y_\mathrm{T})_{ij}|^2 m_\Delta^2 + |\mu|^2 }  \mathrm{ln}\left(1+\frac{m_\Delta^2}{m_{N_k}^2}\right). \label{eq:epsilonD}
\end{align}
For $T > 10^{12}$ GeV, neglecting spectator processes, the Boltzmann equations in these scenarios are given by~\cite{Lavignac:2015gpa}:
\begin{align}
sHz \frac{d\Sigma_\Delta}{dz} &= -\left( \frac{\Sigma_\Delta}{\Sigma_\Delta^{\rm eq}} -1 \right) \gamma_\mathrm{D} 
-2 \left[ \left(\frac{\Sigma_\Delta}{\Sigma_\Delta^{\rm eq}}\right)^2 -1 \right]\gamma_A;\nonumber\\
sHz \frac{d\Delta_\Delta}{dz} &= -\left(  \frac{\Delta_\Delta}{\Sigma_\Delta^{\rm eq}}   -  B_l \frac{Y_\Delta}{Y_l^{\rm eq}} 
+  B_H \frac{Y_\Delta + 2 \Delta_\Delta}{Y_H^{\rm eq}}    \right) \gamma_\mathrm{D};\nonumber\\
sHz \frac{Y_\Delta}{dz} &= -\left( \frac{\Sigma_\Delta}{\Sigma_\Delta^{\rm eq}} -1 \right) \gamma_\mathrm{D} \epsilon_\Delta
-2 B_l \left( \frac{dY_\Delta}{Y_l^{\rm eq}} -\frac{\Delta_{\Delta}}{\Sigma_{\Delta}^{\rm eq}}  \right) \gamma_\mathrm{D}
-2 \left( \frac{Y_\Delta}{Y_l^{\rm eq}} + \frac{Y_\Delta+ 2 \Delta_\Delta}{Y_H^{\rm eq}}  \right) \gamma_{lH},
\label{eq:DBE}
\end{align}
where $\Sigma_\Delta \equiv (n_\Delta + n_{\bar{\Delta}})/s$ and $\Delta_\Delta \equiv (n_\Delta - n_{\bar{\Delta}})/s$ represent the total triplet abundance and asymmetry, respectively, while $B_l \equiv \mathrm{BR}(\Delta \rightarrow \bar{l} \bar{l} )$ and $B_H \equiv \mathrm{BR}(\Delta \rightarrow HH )$ are the branching ratios.

\section{Conditions for $V_\mathrm{R}$ parameters}\label{sec:conditions}
\subsection{Unknown $V_\mathrm{R}$ and constraining conditions}
Despite the fact that the MLRSM model with discrete parity or charge conjugation symmetries allows for the determination of $M_\mathrm{D}$, fully determining $M_\mathrm{D}$ requires knowledge of the heavy neutrino mixing matrix, $V_\mathrm{R}$, which is unknown a priori. 
In the literature, the $V_\mathrm{R}$ matrix is generally taken to be the identity matrix~\cite{Dev:2025fcv}, to coincide with $V_\mathrm{L}$~\cite{Patel:2023voj}, or to be parameterized in relation to $V_\mathrm{L}$~\cite{Li:2024sln,Mikulenko:2024hex}.

In Ref.~\cite{Senjanovic:2019moe}, it is pointed out that the $H$ matrix in Eq.~\eqref{eq:MDH} inherits hermiticity from $M_\mathrm{D}$,  which further implies $\mathrm{Im} \mathrm{Tr} (HH^T)^n=0, (n=1,2,3)$. This serves as a set of constraint conditions for $V_\mathrm{R}$ parameters (and $V_\mathrm{L}$ parameters) when expressed using Eq.~\eqref{eq:HHT}
\begin{align}
\mathrm{Im} \mathrm{Tr}  \left[\frac{v_\mathrm{L}^*}{v_\mathrm{R}}-\frac{1}{M_N} M_\nu^* \right]^n=0,\quad n=1,2,3 
\label{eq:ImTr}
\end{align}
Mathematically, these conditions are equivalent to the requirement that the coefficients of the characteristic polynomial of the $HH^T$ matrix be real, which implies that the eigenvalues of $HH^T$ must be either real or occur in complex conjugate pairs. These conditions correlate the phases of both the left- and right-handed mixing matrices and serve as constraints on the $V_\mathrm{R}$ parameters.

However, these conditions involve a large number of parameters. Even if we fix the dimensionful parameters ($v_\mathrm{L}$, $v_\mathrm{R}$, and the heavy and light neutrino masses), we are left with twelve parameters in the mixing matrices alone. Using the best-fit values for left-handed mixing still leaves eight degrees of freedom. Generally, there are six parameters in $V_\mathrm{R}$: three mixing angles ($\gamma_{ij}$), one Dirac phase, and two Majorana phases. Since there are far fewer equations than parameters, we expect a large degeneracy in the parameter space; namely, for a fixed set of $V_\mathrm{L}$ parameters, there is an infinite set of $V_\mathrm{R}$ parameters satisfying the conditions. Such a situation is certainly daunting for proceeding with the baryogenesis calculation.

Meanwhile, it should be noted that the left-hand sides of the three conditions are exceedingly complicated. Even the simplest condition contains 99 terms, with a total of 2,443 indivisible subexpressions. For the final condition, there are 56,166 terms and 2,018,024 subexpressions. Given this complexity, and the fact that we are primarily interested in the role of low-energy CP-violating parameters, we choose to set all phases in the $V_\mathrm{R}$ sector to zero. This leaves us with three variables, which can then be readily solved using the three equations. In other words, a real $V_\mathrm{R}$ is not merely a convenient choice; only under this condition do we have a fully solvable system (three variables from three conditions).

It should be noted that the constraint conditions follow directly from the hermiticity of $H$, which in turn follows from the hermiticity of $M_\mathrm{D}$. When $\mathcal{C}$ is the LR symmetry, $M_\mathrm{D}$ is symmetric instead of hermitian, leading to no similar constraints. In other words, $V_\mathrm{R}$ is unconstrained in that case. Given the large number of free parameters, we choose to work exclusively with the case where $\mathcal{P}$ is the LR symmetry.

\subsection{Real $V_\mathrm{R}$ and phases of $V_\mathrm{L}$}

The in-principle solvable system—yielding three $V_\mathrm{R}$ mixing angles from three conditions ($\text{Im} \text{Tr} (S^n) = 0$ for $n = 1, 2, 3 ~\text{and}~ S\equiv HH^T$)—is actually difficult to solve given the simultaneous presence of several phases. Even though we work with a real $V_\mathrm{R}$, there are still three phases ($\delta$, $\alpha_{21}$, $\alpha_{31}$) from $V_\mathrm{L}$ present in different parts of the matrix, and the non-linearity grows very quickly because the conditions involve matrix powers. Since we failed to find numerical solutions for the $V_\mathrm{R}$ mixing angles in the presence of all three phases, we instead plot the generated $\text{Im} \text{Tr} (S^n)$ from randomly chosen inputs as a check of existence. This turns out to give us indications about the structure, as detailed below.

\begin{figure}[t!]
\centering
\includegraphics[width=\textwidth]{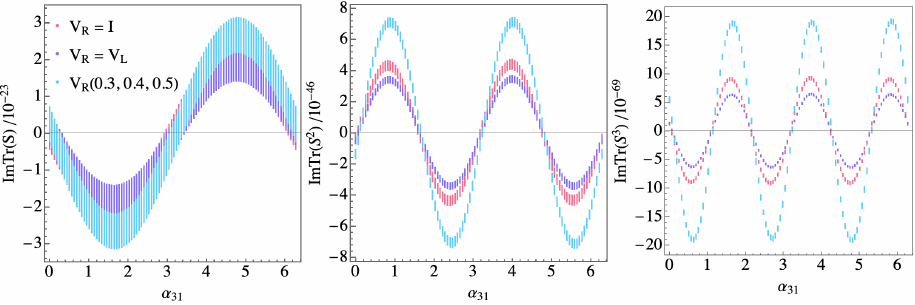}
\caption{Values of $\mathrm{Im} \text{Tr}(S^n)$ for $n=1,2,3$ are calculated using randomly sampled values of $\alpha_{31}$ and $\alpha_{21}$ for three different choices of real $V_\mathrm{R}$. The ``$V_\mathrm{R}=V_\mathrm{L}$" case refers to a real $V_\mathrm{R}$ matrix with mixing angles equal to those of $V_\mathrm{L}$.}
\label{fig:a31}
\end{figure}

\begin{figure}[t!]
\centering
\includegraphics[width=\textwidth]{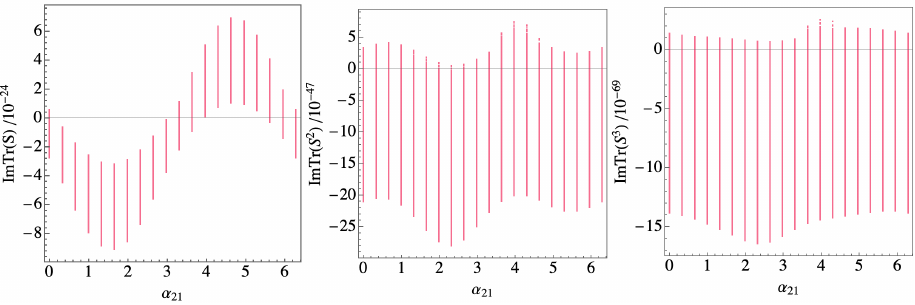}
\caption{Values of $\mathrm{Im} \mathrm{Tr}(S^n)$ for $n=1,2,3$, calculated using randomly sampled values of $\alpha_{21}$ and $V_\mathrm{R}$ mixing angles, with $\alpha_{31}$ fixed to $n\pi$.}
\label{fig:a21}
\end{figure}

First, we want to show that given other parameters, the same values of $\alpha_{ij}$ can simultaneously render all three $\text{Im} \text{Tr} (S^n)$ terms zero. Since a direct numerical solution is not feasible for this problem, we plot the generated $\text{Im} \text{Tr} (S^n)$ instead. The results are shown in Figure~\ref{fig:a31}, where we fix $V_\mathrm{R}$ to be $V_\mathrm{L}$, the identity matrix, or a set of arbitrarily chosen mixing angles, while allowing the two Majorana phases of $V_\mathrm{L}$ to vary in the range $[0, 2\pi]$. The other $V_\mathrm{L}$ parameters are fixed at their best-fit values. The ``width" is purely caused by variations of $\alpha_{21}$. We find that $\mathrm{Im} \mathrm{Tr}(S)$ approaches zero as $\alpha_{31} \to n\pi$. For $\mathrm{Im} \mathrm{Tr}(S^2)$, this occurs at $\alpha_{31} = n\pi/2$, and for $\mathrm{Im} \mathrm{Tr}(S^3)$, at $\alpha_{31} = n\pi/4$. Consequently, to simultaneously satisfy all three conditions, we must choose $\alpha_{31} = n\pi$. No similar behavior was observed for $\alpha_{21}$, which by itself shows that this is a non-trivial result. Considering the daunting amount of subexpressions from the powers of the matrix product, one cannot expect such behavior without computation. We have performed checks with other values of $V_\mathrm{R}$ and confirmed that this is a generic behavior. As a result, we set $\alpha_{31} = n\pi$. It is worth emphasizing that $\alpha_{31} = n\pi$ is not an assumption.

Next, we fix $\alpha_{31} = n\pi$ and vary $\alpha_{21}$ and $\gamma_{ij}$ randomly and plot $\mathrm{Im} \mathrm{Tr}(S^n), n=1,2,3$. The results are shown in Figure~\ref{fig:a21}. Remember that all three $V_\mathrm{R}$ mixing angles are allowed to vary, resulting in a much larger ``width"—the line length for a fixed $\alpha_{21}$. From this figure, we see that while the behavior is less decisive for $\mathrm{Im} \mathrm{Tr}(S^2)$ and $\mathrm{Im} \mathrm{Tr}(S^3)$, $\mathrm{Im} \mathrm{Tr}(S)$ approaches zero when $\alpha_{21}$ takes the CP-conserving values $n\pi$, regardless of the values of $\gamma_{ij}$.

\begin{figure}[t!]
\centering
\includegraphics[width=0.6\textwidth]{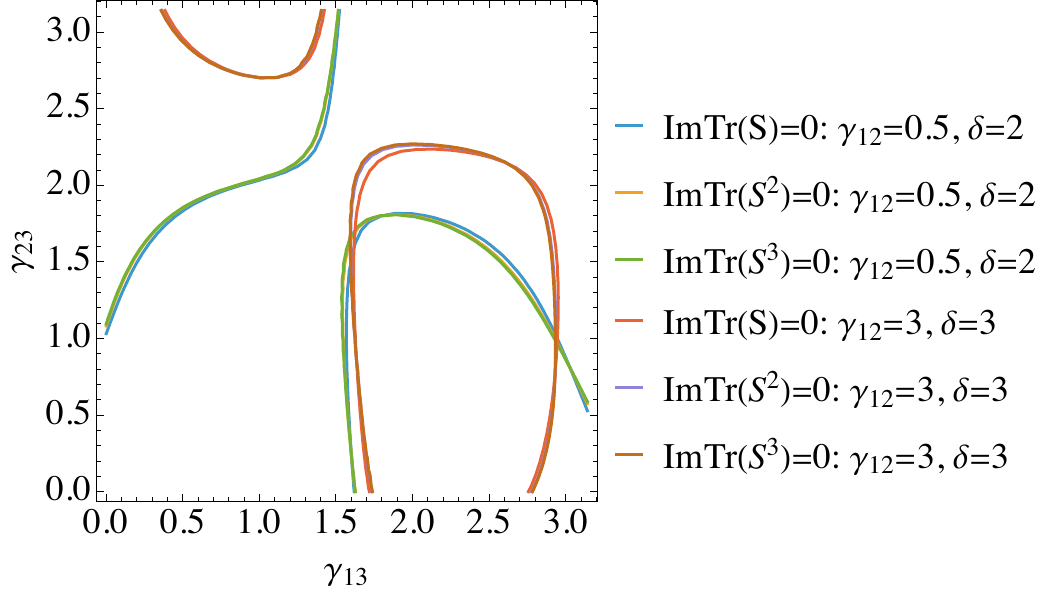}
\caption{The real-trace contours ($\text{Im} \text{Tr} (S^n) = 0$ for $n = 1, 2, 3$) are plotted for two $V_\mathrm{R}$ mixing angles, using two sets of randomly chosen values for $\gamma_{12}$ and $\delta$. The intersection of the three lines (for a given set of inputs) indicates a simultaneous solution to the three constraints.}
\label{fig:only-delta}
\end{figure}

The last thing we want to check is whether a solution exists with both Majorana phases of $V_\mathrm{L}$ set to CP-conserving values. In this case, only the Dirac CP phase is nonzero. The imaginary part of the trace of $HH^T$ arises solely from the presence of this Dirac phase. To illustrate that such a solution exists, we plot the contours of the real trace of $S$ for two sets of randomly chosen values for $\gamma_{12}$ and $\delta$ in Figure~\ref{fig:only-delta}. The intersection of the three lines indicates a simultaneous solution. As shown, in both cases, the three contour lines intersect. It should be emphasized that there is nothing special about the chosen $\gamma_{12}$ and $\delta$ values; other random values also render intersections, implying the existence of a solution (three mixing angles of $V_\mathrm{R}$) from the three conditions ($\text{Im} \text{Tr} (S^n) = 0$ for $n = 1, 2, 3$), as expected. The exist of solution shows that by adjusting the three mixing angles in $V_\mathrm{R}$, one can effectively rotate the generations to find a real trace. 

\begin{figure}[t!]
\centering
\includegraphics[width=0.6\textwidth]{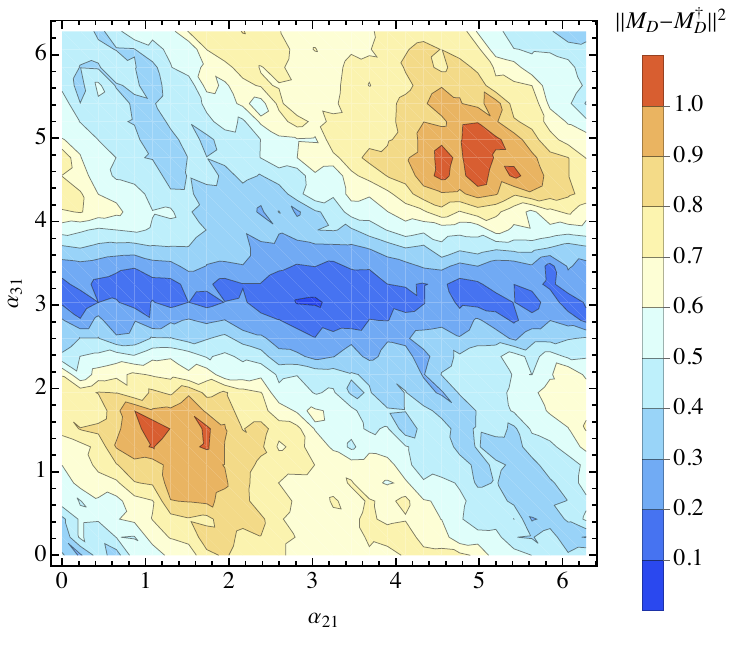}
\caption{Hermiticity check of $M_\text{D}$ against the two $V_\mathrm{L}$ Majorana phases, with $V_\mathrm{R}$ taken real. Other inputs are set for maximal consistency with Figs.~\ref{fig:a31}–\ref{fig:only-delta}. See text for details. }
\label{fig:phaseContour}
\end{figure}

As an independent numerical cross-check, we employ the Casas-Ibarra parametrization for the Dirac mass matrix, $M_\text{D}=i \sqrt{D_N} R \sqrt{\tilde{M}} V_\text{L}^\dagger$, where $D_N$ is the diagonal heavy neutrino mass matrix and $\tilde{M}=D_\nu-V_\mathrm{L}^T M_\mathrm{L} V_\mathrm{L}$ ($D_\nu$ is the diagonal light neutrino mass matrix). We fix the mixing angles of $V_\mathrm{L}$ to their best-fit values, and set $\delta, m_\mathrm{min}, M_1$ to chosen benchmark values. We then jointly optimize over the angles of the complex orthogonal matrix $R$ and the angles of $V_\mathrm{R}$ to minimize the squared Frobenius norm $||M_\mathrm{D}-M_\mathrm{D}^\dagger||^2$ as a function of $\alpha_{21}, \alpha_{31}$. The result is shown in Fig.~\ref{fig:phaseContour}. The hermiticity of $M_\mathrm{D}$ is found to be satisfied around  $\alpha_{21}\simeq \alpha_{31}=\pi$, in full agreement with previous findings.

The results above motivate our choice to work with a real $V_\mathrm{R}$ and vanishing Majorana phases for $V_\mathrm{L}$, and show that solutions exist under these conditions. These results can be understood by considering the different positions of the Dirac and Majorana CP-violating phases. The Dirac phase is inter-generational; in this case, it is easier for the $V_\mathrm{R}$ angles to rotate into a basis with a real trace. In contrast, the Majorana phases are intrinsic to the mass eigenstates. Since a real $V_\mathrm{R}$ can only redistribute the weights of different mass eigenstates, if these weights do not align perfectly to cancel the phases, a real trace cannot be found. It is worth stressing that the CP-conserving tendency of the two $V_\mathrm{L}$ Majorana phases arises when $V_\mathrm{R}$ is real. Once phases are allowed to enter $V_\mathrm{R}$, non-zero phases in $V_\mathrm{L}$ must balance the effect to satisfy the real-trace conditions.

\section{Numeric results}\label{sec:results}

\subsection{Overview of parameters and inputs}
To actually solve for $V_\mathrm{R}$ using the constraints in Eq.~\eqref{eq:ImTr}, we need to specify the values of all other parameters. The model parameters can be categorized as follows:
\begin{itemize}
\item Dimensionful parameters: heavy and light neutrino masses, the triplet mass $m_\Delta$, and the VEVs $v_\mathrm{L}$ and $v_\mathrm{R}$.
\item Dimensionless parameters: mixing angles and phases in both the left- and right-handed mixing matrices.
\end{itemize}

Generally speaking, due to the significantly larger uncertainty in the dimensionful parameters, varying them leads to more pronounced changes in the results. For hierarchical RHNs, the CP asymmetry is proportional to the RHN mass, which leads to the Davidson-Ibarra bound \cite{Davidson:2002qv}. A critical threshold for $m_{N_k}$ occurs when $m_{N_k} < 10^{12}$ GeV (and $10^9$ GeV); at such scales, specific Yukawa interactions enter equilibrium and become distinguishable, meaning flavor effects can no longer be neglected~\cite{Blanchet:2006be}. Another interesting possibility arises when the heavy neutrino mass spectrum is nearly degenerate, leading to a resonant enhancement of the CP asymmetry~\cite{Pilaftsis:2005rv}.
Numerical examples illustrating the effects of varying these dimensionful parameters can be found in Ref. \cite{Zhang:2020lir} (see Figs.~4 and 9 for $m_{N_k}$ and $m_\Delta$, and Fig.~19 for different heavy neutrino mass spectra). From these results, we find that for heavy neutrino mass spectra that are neither degenerate nor highly hierarchical, the impact on the resulting baryon asymmetry is limited. We fix the heavy neutrino mass spectrum to be $m_{N_2}=2 m_{N_1}$ and $m_{N_3}=3 m_{N_1}$, and include the contributions from all RHNs.

The dimensionless parameters serve to redistribute the influence of the dimensionful quantities. This redistribution effect can be quite sharp—potentially driving the CP asymmetries, and consequently the baryon asymmetry, to zero. Fixing the dimensionful parameters allows one to trace the variations of the dimensionless parameters more clearly; in our case, this isolates the effects of the Dirac CP-violating phase. In our numerical study, we fix the left-handed mixing angles to their best-fit values while allowing the Dirac CP-violating phase to vary.

Since this work makes no assumptions regarding the $V_\mathrm{R}$ parameters—which already greatly enlarges the parameter space—we choose to fix the dimensionful parameters to the benchmark values shown in Table~\ref{tab:input}. The $v_\mathrm{L}/v_\mathrm{R}$ values in Scenarios 2 and 4 are obtained from the numerical investigation of type-II mass dominance, as detailed in Appendix A of \cite{Zhang:2020lir}. Once the mass of the decaying particle is chosen, the ratio of the VEVs ($v_\mathrm{L}/v_\mathrm{R}$) is constrained by the mass contribution conditions corresponding to the four scenarios. Fixing these dimensionful parameters allows us to focus exclusively on the effects of Dirac CP violation. Furthermore, as long as the masses do not enter specific regimes (namely the flavored or resonant regimes), one expects similar behavior at other mass scales.

\begin{table}[t!]
 \caption{\label{tab:input} Dimensionful parameters input.}\vspace{.2cm}
 \centering
  \begin{tabular}{c|c|c|c}
   \toprule \hline
    & $m_{N_1}$ [GeV] &  $m_\Delta$ [GeV] & $v_\mathrm{L}/v_\mathrm{R}$\\\hline
  Scenario 1 & $10^{12}$ & $10^{14}$ & $10^{-25}$\\
  Scenario 2 & $10^{12}$ & $5\times 10^{13}$ & $1.45\times 10^{-23}$\\
  Scenario 3 & $10^{14}$ & $5\times 10^{12}$ & $10^{-27}$\\
  Scenario 4 & $10^{14}$ & $5\times 10^{12}$ & $1.45\times 10^{-25}$\\
   \hline
  \bottomrule
\end{tabular}
\end{table}

\subsection{Numeric results and discussions}
\begin{figure}[!ht]
\centering
\includegraphics[width=.92\textwidth]{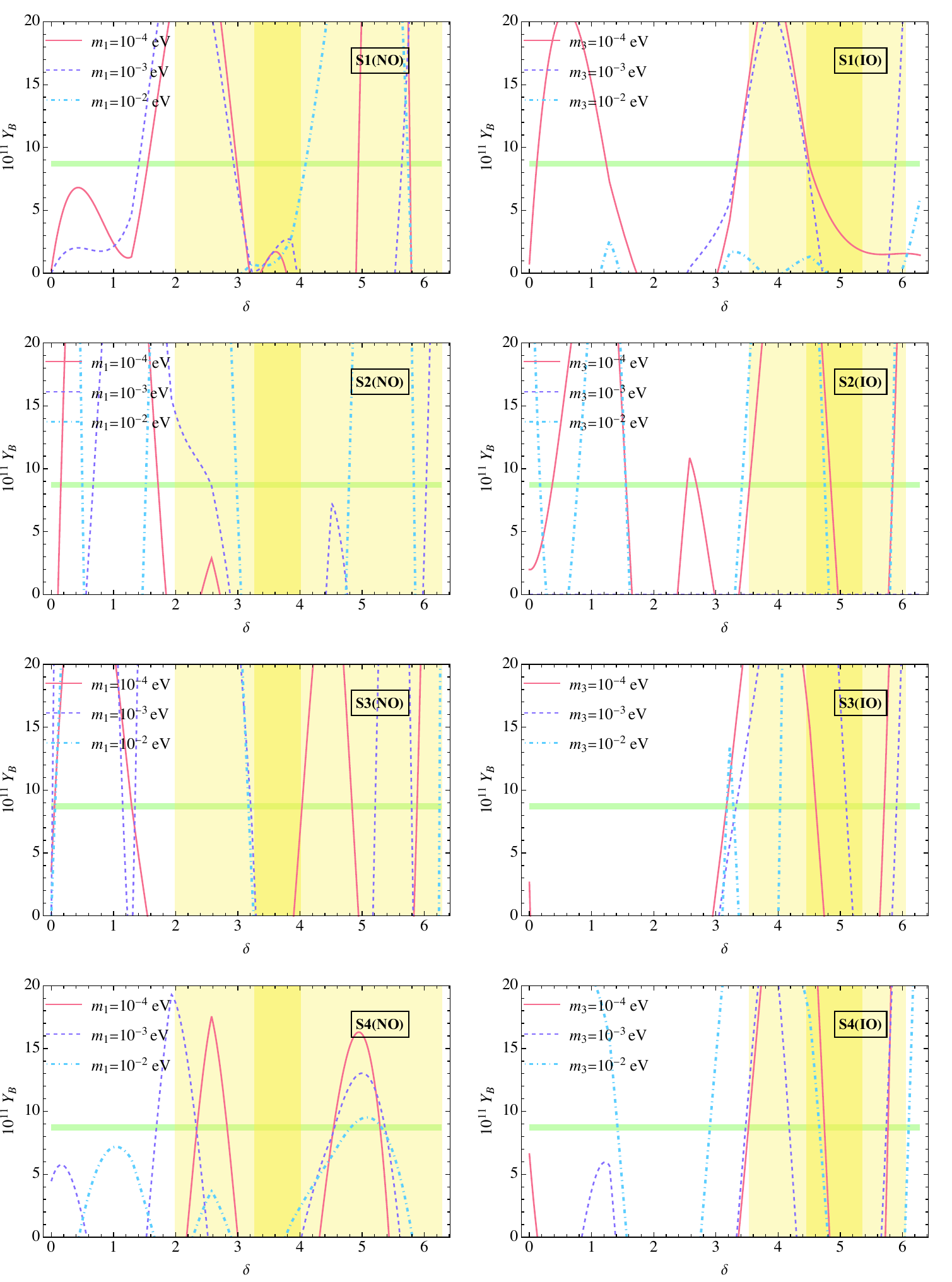}
\caption{Predicted values of the baryon asymmetry as a function of the Dirac CP-violating phase for both mass orderings of light neutrinos in each scenario (labeled S1 through S4). The left panel corresponds to normal ordering (NO), while the right panel corresponds to inverted ordering (IO). In each plot, three benchmark values of the lightest neutrino mass are shown. The green shaded region represents the $3\sigma$ experimental bound on the baryon asymmetry, while the light yellow shaded region indicates the $3\sigma$ range for $\delta$ from NuFIT 6.1~\cite{Esteban:2024eli}, with the darker yellow sub-region representing the $1\sigma$ range.}
\label{fig:yb}
\end{figure}

We select three benchmark values for the lightest neutrino mass:
$m_\mathrm{min}=0.0001, 0.001, 0.01~\mathrm{eV}$,
while sampling the Dirac CP-violating phase evenly in the range $[0, 2\pi]$. The Boltzmann equations in Eqs.~\eqref{eq:NBE} and \eqref{eq:DBE} are employed to calculate the baryon asymmetry for the four scenarios under both light neutrino mass orderings. In Figure~\ref{fig:yb}, we plot the predicted baryon asymmetry as a function of the Dirac CP-violating phase for each scenario. For each input point, we calculate the $V_\mathrm{R}$ parameters from the hermiticity conditions in Eq.~\eqref{eq:ImTr} and then proceed to calculate the baryon asymmetry.

A primary observation from Figure~\ref{fig:yb} is that the predicted curves do not exhibit a simple oscillatory or sinusoidal dependence on $\delta$. This is expected, as each point in the parameter space corresponds to distinct values of the RHN mixing $\gamma_{ij}$ determined through the hermiticity conditions in Eq.~\eqref{eq:ImTr}.

We display only the regions where $Y_\mathrm{B}$ carries the correct sign and show observational constraints on $Y_\mathrm{B}$ and $\delta$ as color-shaded bands. The ``blank" regions—visible, for instance, in scenario S3(IO)—correspond to parameter space points yielding negative $Y_\mathrm{B}$ values. For our chosen benchmark values of $m_\mathrm{min}$, we find several intersections with the experimental $Y_\mathrm{B}$ band where the observed asymmetry is successfully reproduced. While some of these intersections fall outside the $3\sigma$ (or $1\sigma$) confidence levels for $\delta$ (shaded in yellow), every scenario overlaps with at least the $3\sigma$ favored region. This confirms that feasible solutions satisfying both cosmological and oscillation constraints exist across all considered cases.

It is evident that no universal behavior governs these scenarios; the curves vary significantly with $m_\mathrm{min}$, the mass ordering, and the specific leptogenesis scenario. This underscores the high sensitivity of the baryon asymmetry to the lightest neutrino mass and the mass ordering. Notably, the dependence on the Dirac CP-violating phase is remarkably sharp. For a fixed $m_\mathrm{min}$, varying $\delta$ can shift $Y_\mathrm{B}$ by one (e.g., S4(NO)) to two orders of magnitude (e.g., S2(NO) and S3(NO)). Such sensitivity ensures that leptogenesis in these models will provide a stringent test for future high-precision measurements of $\delta$.

Our results provide a consistent mapping from the $(\delta, m_\mathrm{min})$ plane to the final baryon asymmetry for a given scale of the heavy decaying particle, $m_\mathrm{H}$. Although this study was conducted with a fixed $m_\mathrm{H}$ for each scenario, the underlying framework remains general. As future experiments continue to constrain the light neutrino sector, it will be possible to invert this relationship: by utilizing the observed baryon asymmetry as a benchmark, one could potentially probe or constrain the mass scale $m_\mathrm{H}$ itself. This is particularly significant given that these high-energy scales are generally beyond the direct reach of current and planned collider experiments.

\section{Conclusions}\label{sec:conclusions}

Leptogenesis arising from low-energy CP violation provides a compelling bridge between the fundamental properties of neutrinos and the evolution of the early universe. While a direct connection is often obscured by the unknown parameters of the high-energy sector, the MLRSM offers a structured framework to link these scales.

In this work, we have investigated leptogenesis with minimal assumptions by determining the $V_\mathrm{R}$ parameters through conditions enforced by the hermiticity of the neutrino Dirac Yukawa coupling under the $\mathcal{P}$ transformation. We find that these real trace conditions favor CP-conserving values for the Majorana phases, effectively leaving the Dirac CP-violating phase as the only source of the asymmetry. By numerically exploring all four possible leptogenesis scenarios within this framework, we demonstrate that viable leptogenesis is achievable, generating the observed baryon asymmetry with the correct sign.

Our results demonstrate that the final baryon asymmetry is highly sensitive to the Dirac CP-violating phase $\delta$, the neutrino mass ordering, and the absolute mass scale $m_\mathrm{min}$. This sensitivity provides a clear path for testing leptogenesis in the MLRSM against upcoming experimental data. By avoiding ad-hoc assumptions regarding neutrino Yukawa couplings and minimizing free parameters, our approach establishes a robust mapping from the $(m_\mathrm{min}, \delta)$ plane to the observed cosmic matter-antimatter asymmetry for a given high-energy scale $m_\mathrm{H}$.

As future experiments like DUNE, JUNO, and Hyper-K continue to constrain the light neutrino sector, these results will allow us to more precisely evaluate the feasibility of thermal leptogenesis in MLRSM. Furthermore, by anchoring our model to the observed baryon asymmetry, this framework allows us to invert the logic: we can potentially probe the heavy mass scale $m_\mathrm{H}$, which remains otherwise beyond the direct reach of current and planned collider experiments. Ultimately, this work provides a consistent mapping that could eventually narrow the allowed range of the lightest neutrino mass while offering insights into the high-energy completion of the lepton sector.

\section*{Acknowledgements}
XYZ would like to thank Jiang-Hao Yu for suggesting this research in the first place and for early collaboration. The authors would like to thank Gang Li for helpful conversations. This work is supported by National Natural Science Foundation of China under grant No.12305116, Start-up Funds for Young Talents of Hebei University (No.521100223012). XYZ is also supported by the China Scholarship Council (CSC).

%bibliographystyle{JHEP}
%\bibliography{MLRSM}

\subsubsection*{Declaration of generative AI and AI-assisted technologies in the manuscript preparation process}

During the preparation of this work, the authors used Gemini, Claude and deepseek for efficiency and grammar check. The authors reviewed and edited the output as needed and take full responsibility for the content of the published article.

\providecommand{\href}[2]{#2}\begingroup\raggedright\endgroup

\end{document}